\documentclass{emulateapj}

\newcommand{\ldl}{$\lambda/{\Delta}{\lambda}$}
\newcommand{\teff}{T$_{eff}$}
\newcommand{\ki}{\ion{K}{1}}
\newcommand{\csi}{\ion{Cs}{1}}
\newcommand{\nai}{\ion{Na}{1}}
\newcommand{\rbi}{\ion{Rb}{1}}

\newcommand{\ali}{\ion{Al}{1}}

\newcommand{\wat}{H$_2$O}

\newcommand{\hh}{H$_2$}
\newcommand{\kms}{km s$^{-1}$}
\newcommand{\cmss}{cm s$^{-2}$}

\slugcomment{Accepted to ApJ}

\shorttitle{LEHPM 2-59}
\shortauthors{Burgasser \& Kirkpatrick}

\begin{document}

\title{Discovery of the Coolest Extreme Subdwarf}

\author{Adam J.\ Burgasser\altaffilmark{1}}
\affil{Massachusetts Institute of Technology, Kavli Institute for Astrophysics and Space Research,
Building 37, Room 664B, 77 Massachusetts Avenue, Cambridge, MA 02139; ajb@mit.edu}
\and
\author{J.\ Davy Kirkpatrick}
\affil{Infrared
Processing and Analysis Center, M/S 100-22, California Institute
of Technology, Pasadena, CA 91125; davy@ipac.caltech.edu}

\altaffiltext{1}{Visiting Astronomer at the Infrared Telescope Facility, which is operated by
the University of Hawaii under Cooperative Agreement NCC 5-538 with the National Aeronautics
and Space Administration, Office of Space Science, Planetary Astronomy Program.}

\begin{abstract}
We report the discovery of LEHPM 2-59 as the coolest extreme M subdwarf (esdM)
found to date.  Optical and near infrared spectroscopy demonstrate
that this source is of later spectral type than the esdM7 APMPM 0559-2903,
with the presence of strong alkali lines (including \ion{Rb}{1}),
VO absorption at 7400~{\AA} and H$_2$O absorption at 1.4~$\micron$.
Current optical classification schemes yield a spectral type of esdM8,
making LEHPM 2-59 one of only two ultracool esdMs known.
The substantial space velocity of this object ($V_{galactic} \approx -180$ \kms)
identifies it as a halo star.
Spectral model fits to the optical and near infrared spectral data
for this and four other late-type esdMs
indicate that LEHPM 2-59 is the coolest esdM currently known, with
%of NextGen and AMES Cond models. 
%These fits indicate that LEHPM 2-59 has 
{\teff} = 2800-3000 K
and -1.5 $\lesssim$ [M/H] $\lesssim$ -2.0.
Comparison of 
{\teff} determinations for M dwarfs and esdMs based on spectral
model fits from this study 
and the literature demonstrate a divergence 
in {\teff} scales beyond spectral types $\sim$M5/esdM5, 
as large as 600-800~K by types M8/esdM8. While this divergence is likely
an artifact of the underlying classification scheme, it may lead to
systematic errors in the derived properties of intermediate metallicity
subdwarfs. We comment on the future 
of ultracool subdwarf classification, and suggest several ideas
for addressing shortcomings in current (largely extrapolated)
schemes.
\end{abstract}


\section{Introduction}

Subdwarfs are metal deficient stars lying below
%(or more accurately, to the left) of
the stellar main sequence in optical color-magnitude diagrams \citep{kui39}.
Low mass subdwarfs typically exhibit 
halo kinematics (${\langle}V{\rangle} = -202$ km s$^{-1}$; Gizis 1997), and are presumably relics of the early
Galaxy, with ages $\gtrsim$10~Gyr.  With their extremely long lifetimes
(far in excess of the age of the Universe), low mass subdwarfs are important tracers of Galactic structure and chemical enrichment history, 
and are representatives of the first generations of star formation.

The optical and near infrared spectra of the coolest known M-type subdwarfs, like their solar-metallicity
dwarf counterparts, are dominated by molecular absorption, including bands
of CO, TiO, AlH, CaH, CrH, FeH, MgH and {\wat} \citep{mou76,bes82,lie87,giz97,leg98,leg00}.  
Collision induced {\hh} absorption \citep{lin69,sau94,bor97} is also a 
strong absorber around 2 $\micron$ \citep{mou76,leg00}.
Variations in elemental abundances (i.e., metallicity) can modulate
these molecular signatures appreciably, affecting both the total
molecular opacity (and hence overall luminosity at a given mass) 
and relative band strengths through differential
chemical abundance patterns and
modified atmospheric chemistry.
At optical wavelengths, metallicity effects in subdwarf spectra
are seen succinctly 
in the relative bandstrengths of metal oxides and metal hydrides; 
the former are weaker and the latter stronger
in lower metallicity dwarfs \citep{bid76,mou76,cot78,bes82}.  
Current optical classification schemes for M subdwarfs
are generally tied to the relative strengths of these bands.
For example, the most widely used scheme, that defined by \citet[hereafter G97]{giz97},
divides metal-poor M stars into subdwarf (sdM)
and extreme subdwarf (esdM) classes
based on the relative strengths of
CaH and TiO bands in the 6300-7200 {\AA} spectral region.
G97 and \citet{giz97b} have determined mean metallicities of
[Fe/H] = $-1.2\pm$0.3 and [Fe/H] = $-2.0\pm$0.5 for these two classes
of metal-poor dwarfs.

Halo stars exhibit 
large space velocities relative to the Sun; hence, 
they are efficiently detected in proper motion surveys.
Indeed, most low mass subdwarfs now known 
were originally identified in blue photographic plate
proper motion surveys, particularly those by Luyten 
(e.g., LHS and NLTT catalogs; Luyten 1979a,b; see also
Bakos, Sahu \& N{\'{e}}meth 2002 and Salim \& Gould 2003).  
With the digitization of the red optical $R$- and $I$-band
photographic plate
UK Schmidt SERC and AAO \citep{har81,can84,mor92}, 
ESO \citep{wes82,wes84} and Palomar 
(POSS-I, Abell et al.\ 1959; POSS-II, Reid et al.\ 1991) sky surveys,
new proper motion surveys have begun to identify
even cooler objects, which emit more of their light at
wavelengths redward of the visual band.  These
surveys -- including the Automated
Plate Measuring machine Proper
Motion survey \citep[hereafter APMPM]{sch00}, 
the Calan-ESO proper motion survey \citep{rui01}, 
the SUPERBLINK survey \citep{lep02,lep05}
and the SuperCOSMOS Sky Survey \citep[hereafter SSS]{ham01a,ham01b,ham01c} --
and associated follow-up programs\footnote{See
\citet{rey02,lep03,lep0822,lep1425,roj03,por03,por04,ham04,rey04,sch1013,sch1444,dea05,lod05}; and \citet{sub05a,sub05b}.}
have pushed subdwarf discoveries to the end of the
M spectral class \citep{sch1013}.  Metal-poor
analogs to even cooler L dwarfs \citep{kir99} have been identified
in the SUPERBLINK survey \citep[hereafter LRS03]{lep03} and serendipitously
\citep{me03f,me04} in the Two Micron All Sky Survey \citep[hereafter 2MASS]{skr06}.
These low temperature {\em ultracool subdwarfs},
encompassing objects with spectral types sdM7/esdM7 and later 
\citep[see also Kirkpatrick, Henry \& Irwin 1997]{mecs13}
provide new challenges for atmospheric modeling
and extend our sampling of the halo population down to and below
the hydrogen burning minimum mass \citep{me03f}.

While several ultracool subdwarfs are now known to exist,
only one ultracool extreme subdwarf has been found, the esdM7 
APMPM 0559-2903 \citep{sch99}.
To seek out even cooler subdwarfs, 
we have initiated a program to obtain spectra of red proper
motion stars selected from the Liverpool Edinburgh
High Proper Motion survey \citep[hereafter LEHPM]{por03,por04}.
One of these sources, 
LEHPM 2-59, appears to be a 
later-type esdM than APMPM 0559-2903 based on both near infrared and optical
data, and we identify it as the latest-type and coolest
esdM found to date.
In $\S$2 we summarize the selection of this source from 
the LEHPM survey.  In $\S$3 we describe near infrared
and optical spectroscopic observations
of this and four other late-type esdMs, and describe their
overall characteristics.  In $\S$4 we analyze
line strengths, radial velocities and spectral types 
for the optical data, and estimate the distance and kinematics of LEHPM 2-59.
In $\S$5 we use spectral model fits to the observed
optical and near infrared data
to derive {\teff}s and metallicities
for the latest-type esdMs. 
We discuss these results in $\S$6, focusing on
the temperature scale of esdMs and future revisions to existing
classification schemes for ultracool subdwarfs.  Results are summarized in $\S$7.

\section{Selection of LEHPM 2-59}

The LEHPM catalog is a subset
of 11289 proper motion stars from the SSS detected 
in $R$-band ESO and UK Schmidt plates covering 7000 deg$^2$ of
the Southern sky ($\delta \lesssim -20\degr$).  This area 
excludes regions close to the Galactic plane and those
fields with epoch differences less than 3 yr (the mean
epoch difference is 8.5 yr).
Catalog sources were selected to have
0$\farcs$18 yr$^{-1}$ $\leq \mu \leq$ 20$\farcs$0 yr$^{-1}$,
${\mu}/{\sigma}_{\mu} > 3$ 
and 9 $\leq R \leq$ 19.5, and were cross-matched with 
photographic $B_J$ and $I_N$ plates (via the SSS) and
the 2MASS point source catalog.
Further details on the construction and completeness of the LEHPM
catalog are given in \citet{por04}.

We selected ultracool dwarf and subdwarf 
candidates from the LEHPM catalog based on 
red optical/near-infrared color and $J$-band reduced proper motion 
(RPM; Hertzsprung 1905; Luyten 1922), 
$H_J \equiv J + 5\log_{10}{\mu} + 5 = M_J + 5\log_{10}{V_{tan}} - 3.73$.
RPM provides a measure of an object's absolute brightness
and tangential space velocity ($V_{tan}$) independent of its 
(generally unknown) distance,
and therefore enables the identification of intrinsically faint
and/or halo stars. Optical/near infrared RPM diagrams have become increasingly
popular in searches for ultracool nearby and halo stars \citep{sal02}, 
as these low temperature sources emit
more of their flux outside of the optical bands. Here, we focus 
on the $J$-band 
RPM since the spectral energy distributions of late-type dwarfs 
peak at these wavelengths.
Figure~\ref{fig_rpm} displays the $H_J$ versus $(R_{ESO}-J)$ diagram for
the LEHPM catalog.  The main cluster of sources running diagonally
down the middle of the diagram is composed primarily of main sequence dwarfs.
The smaller cluster of sources offset below and to the left is composed
of metal-poor (shifting their colors toward the blue), high velocity
(increasing $H_J$) halo subdwarfs.  The smallest grouping
in the lower left corner of the diagram is primarily composed of low
luminosity white dwarfs.

To select for the latest-type sdMs and esdMs, we applied the following selection criteria
to the LEHPM catalog:
\begin{itemize}
\item Detection in both $R_{ESO}$ and $J$ bands, and
\item $(R_{ESO}-J) \ge 1.5$ and $H_J \ge 19.25$ {\em or} 
\item $(R_{ESO}-J) \ge 3.5$ and $H_J \ge 24.5 - 1.5(R_{ESO}-J)$.
\end{itemize}
The color/RPM criteria are illustrated in Figure~\ref{fig_rpm}. 
Imposing $(R_{ESO}-J) \ge 1.5$ eliminates contamination by the (relatively few)
cool white dwarfs; the remaining criteria cordon off the late-type 
extensions of the dwarf and subdwarf tracks.  A total of 50
sources were selected in this manner, including LEHPM 2-59.  Of these,
14 have been previously observed by other programs, including
7 M8-M9 dwarfs \citep{giz02,lod05}, 
the sdM7 SSSPM J1930-4311 \citep[a.k.a.\ LEHPM 2-31]{sch1013} and the 
esdM6 SSSPM J0500-5406 \citep[a.k.a.\ LEHPM 1-3861]{lod05}.  Results for our complete
LEHPM sub-sample will be presented in a forthcoming publication.

\section{Observations}

\subsection{Near Infrared Spectroscopy}

LEHPM 2-59 and the three late-type esdMs 
LP 589-7 \citep[esdM5]{giz99}, 
LSR 0822+1700 \citep[esdM6.5]{lep0822} and
APMPM 0559-2903 \citep[esdM7]{sch99} were each
observed with the SpeX spectrograph \citep{ray03}, 
mounted on the 3m Infrared Telescope Facility, during three runs
on 2004 March 10, 2004 September 5--9 and 2005 December 31 (UT).
A log of observations is provided in Table~\ref{tab_spexlog}.
Conditions during these runs ranged from poor (cloudy and high humidity)
during 2004 March, to clear with light cirrus during the 2004 September and
2005 December runs.  Seeing was typically 0$\farcs$5--0$\farcs$9 on all
nights.  Spectral data were
obtained using the prism dispersed mode, which
provides low resolution 0.7--2.5~$\micron$ spectra in a single order.
Using a 0$\farcs$5 slit, we obtained data with
spectral resolution {\ldl} $\approx 150$
and dispersion across the chip of 20--30~{\AA}~pixel$^{-1}$.
The slit was oriented to the parallactic angle in all observations to reduce
differential color refraction, and the telescope was
guided by spillover light from the targets in the imaging channel.
Multiple exposures of 180~s each were obtained 
in an ABBA dither pattern along the slit.
Nearby A0~V stars were observed immediately
after the target observations and at a similar airmass (${\Delta}z < 0.1$) 
for flux calibration and telluric corrections.
Internal flat field and Ar arc lamps were also observed with each target
for pixel response and wavelength calibration.

All spectral data were reduced using the SpeXtool package, version 3.3
\citep{cus04} using standard settings.
First, spectral images were corrected for linearity, pair-wise subtracted, and divided by the
corresponding median-combined flat field image.  
Spectra were then optimally extracted using
default settings for aperture and background source regions, and wavelength calibration
was determined from arc lamp and sky emission lines. Individual 
spectral observations for each science target and A0 star 
were normalized and
combined using a robust weighted mean with 8$\sigma$
outlier rejection.   Data for the A0 stars were also corrected for broad-band
shape variations before combining.
Telluric and instrumental response corrections for the science data were determined using the method outlined
in \citet{vac03}, with line shape kernels derived from
arc lines.  Adjustments were made to the telluric spectra to compensate
for differing \ion{H}{1} line strengths in the observed A0~V spectrum
and pseudo-velocity shifts.
Final calibration was made by
multiplying the observed target spectrum by the telluric correction spectrum,
which includes instrumental response correction through the ratio of the observed A0~V spectrum
to a scaled, shifted and deconvolved Kurucz\footnote{\url{http://kurucz.harvard.edu/stars.html}.}
model spectrum of Vega.

The reduced near infrared spectra of the four esdMs are shown in Figure~\ref{fig2}.
These spectra exhibit the characteristic
signatures of late-type
esdMs, including blue near infrared spectral slopes
due to strong collision-induced H$_2$ absorption; 
shallow H$_2$O absorption at 1.4 and 1.9 $\micron$; FeH and CrH absorption
at 0.86, 0.99 and 1.2 $\micron$; weak bands of TiO at 0.84 $\micron$;
and \ion{K}{1} and \ion{Na}{1} lines at 0.77, 0.82 and 1.17 $\micron$
\citep{mou76,leg00}.  Notably absent
are the deep H$_2$O and 2.3 $\micron$ CO bands that characterize solar-metallicity 
late-type M dwarfs \citep[cf.\ Fig.\ 1 of Burgasser et al.\ 2004]{bal73,jon94}.
The weakness of these bands makes near infrared classification
of esdMs difficult (Burgasser et al., in prep.; see $\S$6.2);
however, the overall similarity of the spectrum of LEHPM 2-59 to 
those of LSR 0822+1700 and APMPM 0559-2903, 
coupled with its stronger H$_2$O absorption
and bluer spectral slope, strongly suggests a late esdM spectral type.

\subsection{Optical Spectroscopy}

We obtained optical spectroscopy for LEHPM 2-59 and the three
late-type esdMs LP 589-7, SSSPM 0500-5406 
and APMPM 0559-2903 on
2005 December 4 (UT) using the Low Dispersion Survey Spectrograph (LDSS-3)
mounted on the Magellan 6.5m Clay Telescope.  A log of observations
is given in Table~\ref{tab_ldss3log}.
LDSS-3 is an imaging spectrograph, upgraded by M.\ Gladders
from the original
LDSS-2 \citep{all94} for improved
red sensitivity.  The instrument is composed of an
STA0500A 4K$\times$4K CCD camera that re-images an
8$\farcm$3 diameter field of view
at a pixel scale of 0$\farcs$189.  A set of slit masks and grisms
allow spectral observations at various resolutions across the optical
and red optical band.
For our observations, 
we employed the VPH-red grism (660 lines/mm) with a 0$\farcs$75
(4 pixels) wide longslit to
obtain 6050--10500 {\AA} spectra across the entire chip
with an average resolution {\ldl} $\approx$ 1800.  Dispersion
along the chip was 1.2 {\AA}/pixel.  The OG590 longpass filter
was used to eliminate second order light shortward of 6000 {\AA}.
Two long exposures were obtained for each target, followed 
immediately by a series of NeArHe arc lamp and flat-field
quartz lamp exposures. We also observed a nearby 
G2-G3 V star after each esdM target for telluric absorption correction.

Data were reduced in the IRAF\footnote{IRAF is distributed by the National Optical
Astronomy Observatories, which are operated by the Association of
Universities for Research in Astronomy, Inc., under cooperative
agreement with the National Science Foundation.} environment.
We focus our analysis on the blue side of LDSS-3
(in two-amplifier readout mode), covering
the 6050--8400 {\AA} region.
Raw science images
were trimmed and subtracted by a median combined
set of bias frames taken during the afternoon.  
The resulting images were divided by the corresponding 
normalized, median-combined and bias-subtracted set of flat field frames.
Spectra were then
extracted using the APALL task with background subtraction but without
variance weighting (i.e., not ``optimal extraction'').
The dispersion solution for each target 
was determined using the tasks REFSPEC, IDENTIFY and DISPCOR,
and arc lamp spectra extracted
using the same dispersion trace; solutions were typically accurate
to 0.05 pixels, or 0.07 {\AA}.  Flux calibration was determined
using the tasks STANDARD and SENSFUNC with
observations of the spectral standard Hiltner 600 
\citep[a.k.a.\ HD 289002]{ham94}
obtained on 2005 December 3 (UT) with the same slit and grism combination
as the science data.  Corrections to telluric O$_2$ (6850--6900 {\AA} B-band,
7575--7700 {\AA} A-band)
and H$_2$O (7150--7300, 8150--8350 {\AA}) absorption
for each esdM/G star pair
were determined by linearly interpolating over these features in the 
G star spectrum, dividing by the uncorrected G star spectrum, and multiplying the
result with the esdM spectrum.  These corrections were generally adequate,
but noticeable 
residuals are seen around the O$_2$ A-band in the spectra of
SSSPM 0500-5406 and APMPM 0559-2903, for which G star telluric calibrators
were observed at a larger differential airmass.  These residuals 
principally affect
the blue wing of the \ion{K}{1} doublet at 7665/7699~{\AA}.

Reduced spectra for the four esdMs observed are shown in Figure~\ref{fig3}.
Each spectrum shows characteristic spectral traits of late-type
esdMs, including weak TiO absorption bands at 6400, 6700 and 7150 {\AA} 
(and possibly weak TiO absorption at 7800 {\AA}); strong CaH bands
at 6400 and 6900 {\AA}; and prominent line absorption from
\ion{K}{1} (7665/7699 {\AA} doublet) and \ion{Na}{1} (8183/8195 {\AA} doublet).
The \ion{K}{1} doublet lines show substantial broadening in the spectrum of 
LEHPM 2-59, reminiscent of the {\ki} 
broadening observed in solar metallicity L dwarfs.  
The 6708~{\AA} \ion{Li}{1} line is not detected in any of the spectra;
this species has almost certainly been depleted by fusion
reactions in the cores of all of these objects.
There is a dip in the spectra of SSSPM 0500-5406,
APMPM 0559-2903 and LEHPM 2-59 at 7400 {\AA} which we attribute to 
the 1-0 B$^4{\Pi}$-X$^4{\Sigma}^-$ VO band, a feature commonly observed 
in solar-metallicity M dwarf and early L dwarf
spectra but identified here for the first time
in esdM spectra.  This detection is robust despite poor telluric
O$_2$ correction in the spectra of SSSPM 0500-5406 and APMPM 0559-2903, 
as this calibration error affects only the 7600--7700~{\AA}
window, while a dip in the spectrum is seen clearly at 7400--7500~{\AA} (see also $\S$~5).
Using the Kurucz Atomic Line Database\footnote{Obtained
through the online database search form created by C.\ Heise and maintained
by P.\ Smith; see 
\url{http://cfa-www.harvard.edu/amdata/ampdata/kurucz23/sekur.html}.} 
\citep{kur95}, 
we have also identified a forest of \ion{Ca}{1} lines 
between 6100--6170 and 6440--6500 {\AA},
as well as more prominent lines at 6573 and 7326 {\AA}.  
All of the atomic
features identified in these spectra are listed in Table~\ref{tab_atomic};
molecular line identifications can be found in \citet{kir91}. 

{\rbi} lines at 7800 and 7948 {\AA} are also present in the spectra
of APMPM 0559-2903 and LEHPM 2-59, and the 7948 {\AA} line is
weakly present in the spectra of LP 589-7 and SSSPM 0500-5406.  
These lines are particularly interesting, as they have
been previously observed only in the low resolution 
spectra of L dwarfs \citep{kir99}
and, more recently, of the esdM6.5 LSR 0822-1700 \citep{lep0822}
and the L subdwarf LSR 1610-0040 
\citep[however, see Cushing \& Vacca 2006; Reiners \& Basri 2006]{lep1610}. 
On the other hand, {\rbi} is not
detected in low resolution spectra of 
the latest-type M dwarfs or sdMs \citep{kir99,lep1425,sch1013}.
This atomic species is largely ionized in the photospheres of M dwarfs
(T$_{ion}$ $\sim$ 2350 K; Lodders 1999, 2002),
although a significant fraction of {\rbi} gas is probably neutral 
at higher temperatures in the higher pressure
photospheres of subdwarfs.  
The appearance of both the 7400 {\AA} VO band and the
7800/7948 {\AA} {\rbi} doublet lines could be 
considered a defining trait of ultracool esdMs.

Finally, we note that H$\alpha$ emission, a feature 
arising from magnetic activity that is nearly
always present in the spectra of late-type M dwarfs
\citep{giz00,haw02,wes04}, is not detected 
in any of these spectra (including LP 589-7; Gizis \& Reid 1999), 
consistent with weak or absent chromospheres.

\section{Analysis}

\subsection{Atomic Line Strengths and Radial Velocities}

Equivalent widths (EWs) for the alkali lines and the two most prominent
\ion{Ca}{1} lines were measured using the IRAF SPLOT routine.
Uncertainties were determined as the scatter of values from multiple
measurements, while upper limits for non-detected lines ({\rbi} in 
LP 589-7 and SSSPM 0500-5406) were estimated from the mean EWs of
local noise features.
These values are given in Table~\ref{tab_ews}.
Note that EWs given for the \ion{K}{1} and \ion{Na}{1} doublets
are the combined absorption from both lines in each feature.
Nearly all of these lines are seen to strengthen 
from LP 589-7 to LEHPM 2-59, with the possible exception
of the 7326 {\AA} \ion{Ca}{1} line, a transition that has the highest 
lower energy state in Table~\ref{tab_atomic} 
(4.6 eV; Kurucz \& Peytremann 1988).  The strengthening of atomic lines
is
observed in the spectra of solar-metallicity M and L dwarfs as one progresses
to later spectral types, and is attributed to lower photospheric
temperatures and the corresponding increase in column abundances 
of these neutral species.  The evolution of the line strengths
in the esdMs observed here also suggests 
a trend of cooler {\teff} with later esdM type.

The six alkali lines and two strongest \ion{Ca}{1} lines were used to 
measure radial velocities ($V_{rad}$)
for the observed sources. 
Line centers were determined
by gaussian fits to the line cores, and velocity shifts 
were determined relative to the vacuum wavelengths listed
in the Kurucz database (Table~\ref{tab_atomic}).  
The mean and standard deviations of these shifts are listed in Table~\ref{tab_ews},
and include a systematic uncertainty of 3 {\kms} based on the mean
accuracy of the dispersion solutions.  
Our values for LP 589-7, SSSPM 0500-5406 and APMPM 0559-2903
are in agreement with those of \citet{giz99}; \citet{lod05}; and \citet{sch99},
respectively.  With $V_{rad}$ = 79$\pm$9 \kms, LEHPM 2-59 also appears to 
have kinematics consistent with a halo subdwarf.

\subsection{Spectral Classification}

The optical spectra were classified using the scheme of G97 as
updated by LRS03.  This scheme is based on the relative strengths
of the 7100 {\AA} TiO band and the 6400/6900 {\AA} CaH bands,
as sampled by the four indices CaH1, CaH2, CaH3 and TiO5 defined
in \citet{rei95}; as well as the redness of the pseudo-continuum
in the 6500--8200 {\AA} spectral region as sampled by the Color-M
index of LRS03.  
These indices were measured for each of the spectra after shifting
them to their rest frame velocities.  Figure~\ref{fig4}
compares the combined CaH2+CaH3 indices to TiO5 for these
sources, in addition to measurements for late-type dwarfs
from \citet{haw96}; G97; \citet{giz97b};
\citet{rei05}; LRS03; \citet{lep0822}; and \citet{sch1013}. 
This index-index diagram, based on work by G97, 
has been used by several groups to segregate M
dwarf spectra by metallicity class.
We update the divisions originally set forth by G97 for the combined
CaH2+CaH3 indices:
%\begin{equation}
%sdM: ({\rm CaH2+CaH3}) < 1.31({\rm TiO5})^3 - 2.37({\rm TiO5})^2 + 2.66({\rm TiO5}) - 0.20 \label{eqn_dsd}
%\end{equation}
\begin{eqnarray}
sdM: & ({\rm CaH2+CaH3}) < 1.31({\rm TiO5})^3 - \nonumber \\
 & 2.37({\rm TiO5})^2 + 2.66({\rm TiO5}) - 0.20 \label{eqn_dsd}
\end{eqnarray}
%\begin{equation}
%esdM: ({\rm CaH2+CaH3}) < 3.54({\rm TiO5})^3 - 5.94({\rm TiO5})^2 + 5.18({\rm TiO5}) - 1.03. \label{eqn_sdesd}
%\end{equation}
\begin{eqnarray}
esdM: & ({\rm CaH2+CaH3}) < 3.54({\rm TiO5})^3 -  \nonumber \\
 & 5.94({\rm TiO5})^2 + 5.18({\rm TiO5}) - 1.03. \label{eqn_sdesd}
\end{eqnarray}
We stress that, as in G97, these divisions are somewhat arbitrary as
sources spanning Figure~\ref{fig4} represent a broad continuum
of metallicities.  Figure~\ref{fig4} nevertheless 
demonstrates that SSSPM 0500-5409, APMPM 0559-2903 and LEHPM 2-59
are all bona-fide esdMs lying at the tail end of the
esdM distribution.  LP 589-7 appears to be a borderline sd/esd object;
however, for the remainder of this paper we will treat it as an esdM.

Numerical subtypes for the four observed 
sources were determined using the relations
\begin{equation}
{\rm SpT}_{esdM} = 7.91({\rm CaH2})^2 - 20.63({\rm CaH2}) + 10.71 \\
\end{equation}
\begin{equation}
{\rm SpT}_{esdM} = -13.47({\rm CaH3}) + 11.50 \\
\end{equation}
(i.e., Eqns.\ 2 and 8 from G97), 
and 
\begin{equation}
SpT_{esdM} = 22.0\ln({\rm Color-M}) - 4.1
\end{equation}
(i.e., Eqn.\ 14\footnote{Note that the LRS03 
Color-M/spectral type relation uses the natural logarithm of the Color-M index, 
and not the base-10 logarithm as suggested by their Eqn.\ 14.} 
in LRS03),
where SpT$_{esdM}$(esdM0) = 0, SpT$_{esdM}$(esdM5) = 5, etc.
Technically, these relations are defined only up to subtype esdM6 (G97), but we
follow current practice \citep{lep1425,lep0822,sch1013,lod05} 
in extrapolating these relations to later numerical types (however, see $\S$6.2).
Index values, index subtypes and averages of the numerical subtypes 
(rounded off to the nearest 0.5 subtype) are listed
in Table~\ref{tab_class}.  

The derived subtypes for LP 589-7
and APMPM 0559-2903 are consistent with
previous determinations.  We derive an esdM6.5 classification
for SSSPM 0500-5406, 0.5 subtypes later than the classification
given by \citet{lod05} who measured  
CaH2 and CaH3 indices 0.05 larger than our values.  However, this is 
generally consistent
with the typical uncertainties in index measurements and 
the corresponding uncertainties in derived spectral 
types ($\pm$0.5 subtype).  For LEHPM 2-59 we derive a 
classification of esdM8, a full
subtype later than APMPM 0559-2903, until now the
latest esdM known.  This result 
is consistent with the observed spectral trends noted above,
particularly the deeper near infrared {\wat} bands 
and stronger optical atomic lines in the spectrum of 
LEHPM 2-59.
We conclude that LEHPM 2-59 is the latest-type esdM 
found to date.

\subsection{Distance and Space Velocity Estimates}

There are currently few late-type esdMs with measured parallaxes;
only two sources esdM5 and later (the esdM5 LHS 3061 and the esdM6 LHS 1742a;
Monet et al.\ 1992) have known
distances.  Hence, only a rough distance estimate ($d_{est}$) for LEHPM 2-59 is 
possible.  For this, we employed the spectral type/absolute magnitude relations of LRS03 (Eqns.\ 24 and 25), which can be rewritten as
\begin{equation}
M_R = 10.44 + 0.49({\rm SpT}_{esdM}) \\
\end{equation}
\begin{equation}
M_{K_s} = 8.06 + 0.37({\rm SpT}_{esdM}), \\
\end{equation}
where $R$ is from USNO-B1.0 \citep{mon03}
and $K_s$ is from 2MASS.  \citet{rei05} derived a similar relation for
$M_{K_s}$ versus spectral type for esdMs,
\begin{equation}
M_{K_s} = 8.73 + 0.31({\rm SpT}_{esdM}). \\
\end{equation}
Assuming SpT$_{esdM}$ = 8,
$R$ = 18.82 and $K_s$ = 14.76 (Table~\ref{tab_properties}),
these three relations yield distances of 79, 55 and 51 pc,
respectively.  \citet{giz99}
have also derived absolute magnitude/color relations for esdMs; in particular,
\begin{equation}
M_{I} = 5.96 + 4.29(R-I), \\
\end{equation}
where $R$ and $I$ are photographic magnitudes.  Again, adopting
photometry from USNO-B1.0 ($R$ = 19.92, $I$ = 17.24)
yields $d_{est}$ = 79 pc, in agreement with the $M_R$/spectral
type relation in Eqn.\ 6.  We therefore adopt an average
$d_{est}$ = 66$\pm$15~pc for this source, 
but stress that such estimates are highly
uncertain given the current paucity of parallaxes 
for late-type subdwarfs. 
% Interestingly, 
%LEHPM 2-59 has a similar distance
%estimate as APMPM 0559-2903 \citep{lep0822}.

Using this estimated distance, and the measured proper motion
and radial velocity of LEHPM 2-59, we computed
$UVW$ space velocity components relative to the Local Standard of Rest (LSR).
We derive [$U,V,W$] $\approx$ [135,-180,-80] {\kms}, assuming an LSR solar motion 
of [$U_{\sun},V_{\sun},W_{\sun}$] = [10,5,7] {\kms} \citep{deh98}.  
These velocities lie well
outside of the velocity dispersion sphere of local disk M dwarfs 
([$\sigma_U$,$\sigma_V$,$\sigma_W$] $\approx$ [40,28,19] {\kms}
centered at [-13,-23,-7] {\kms}; Hawley, Gizis \& Reid 1996);
and the large negative $V$ velocity, ranging 
over -210 to -145 {\kms} for our distance estimates,
is consistent with motion independent of the Galactic disk.  
Assuming an LSR rest frame velocity of 
-220 {\kms} \citep{ker86}, the total space 
velocity of LEHPM 2-59 in the Galactic potential is
135--190 \kms.  This is well below the Galactic escape velocity
in the vicinity of the Sun \citep{car88,leo90}. % [ORBIT?]
Interestingly, the orbit of this object is highly elliptical,
and passes well inside the Galactic bulge on its closest
approach to the Galactic center.
This is not atypical for halo stars currently in the vicinity of the Sun
(S.\ L\'epine, 2006, private communication).
 
\section{Spectral Model Fits}

To further gauge the physical properties of these late-type esdMs, we 
compared our optical and near infrared spectra to subsolar metallicity
theoretical spectral models from \citet[NextGen]{hau99}
and \citet[AMES Cond]{all01}. These models are based on the
Phoenix code \citep[and references therein]{hau97}, 
employ self-consistent temperature/pressure profiles, and assume
local thermodynamic equilibrium.  Both sets of models use 
the opacity sampling method; a full account of the chemical species and
opacities used by these models is provided in 
\citet{all01} and references therein.
Two major changes in the Cond models are
the removal of condensed species from the upper atmosphere
and the use of updated TiO and H$_2$O opacities.
The NextGen models and their antecedents \citep{all90,all95,all97}
have been used extensively for
fitting M subdwarf spectra \citep{giz97,sch99,daw00,leg00,lep0822}.

We sampled grids of NextGen and Cond 
models spanning temperatures of 2600 $\leq$ \teff $\leq$ 3600~K
in steps of 100~K, metallicities of
-3.0 $\leq$ [M/H] $\leq$ 0.0~dex in steps of 0.5~dex, 
and surface gravities of
$\log{g}$ = 5.0 and 5.5 \cmss.  Comparisons were made
separately to the optical and near infrared data.  For the optical fits,
both empirical and model spectra
were normalized at 8100~{\AA} as in Figure~\ref{fig3}, and the observed
data were shifted to their rest frame velocities.
For the near infrared fits, data and models were scaled to 1.2 $\micron$.
Model spectra were also deconvolved to the resolution of
the observed data ({\ldl} = 1800 and 150 for the optical and near infrared,
respectively) using a Gaussian kernel.
For each spectrum/model pairing,
the root mean square (RMS) deviation was computed over the spectral
ranges 6050--7500 and 7650--8250 {\AA} in the optical (to avoid regions of poor
telluric O$_2$ correction and the 7400 {\AA} VO band; see below);
and 0.75--1.31, 1.45--1.75 and 2.0--2.4 $\micron$ in the near infrared
(to avoid telluric H$_2$O regions).
The normalization of the 
model spectra in each fit was allowed to vary by a factor ranging from 0.75
to 1.25 (in steps
of 0.05) to allow for continuum offsets, and the normalization
with the minimum RMS was retained.  For each
model set and surface gravity, the {\teff} and [M/H] combination
with the overall minimum RMS 
was deemed to be the best fit
for a particular spectrum.  
%In all cases, these fits converged to a single best model (i.e., there
%were no local minima), although there was some degeneracy between 
%[low, high] {\teff} and [low, high] [M/H].
These best fit values are listed in Table~\ref{tab_model}.

Figure~\ref{fig5} and~\ref{fig6} 
compares the best fit NextGen and Cond, 
$\log{g} = 5.0$ models for the spectrum of LEHPM 2-59 in the optical and near infrared,
respectively.  Overall, these fits are reasonably
good, with the broad spectral shape and several individual features matching
quite well.  For the optical data, the $\log{g} = 5.0$ NextGen models
consistently provided the best fits to the data.  This appears
to be largely due to better fits to 
the 6700~{\AA} CaH band which is only slightly overestimated
in the NextGen models but greatly
underestimated in the Cond models.  The
7150 {\AA} TiO band is too strong in both model sets; 
higher {\teff} and/or lower [M/H] models 
weaken this band but show very poor agreement with
the rest of the optical spectrum.  TiO features have long been identified as
a problem, particularly in the NextGen models \citep{all00}.
Atomic lines are also generally too strong in the best fit models.  
Indeed, there are many metal lines apparent in the Cond
models for which no analogs are seen in the empirical data.  
The \ion{Na}{1}, \ion{Rb}{1} and prominent \ion{Ca}{1}
lines are also too deep, although 
the \ion{K}{1} doublet is reproduced quite well, both in depth
and breadth, for both sets of models.
The apparent mismatch over 7300-7500 {\AA} is due to the 7400 {\AA} VO band, which
is not included in the opacity set of either the NextGen 
or Cond models \citep{all01}.  We found
slight systematic effects in the derived parameters between different 
model sets and surface gravities.  The best fit Cond models 
yielded slightly lower {\teff}s (by $\lesssim$ 100 K) and higher
[M/H] values (by $\lesssim$ 0.5 dex) as compared to the best fit NextGen models;
higher gravity model fits typically gave higher {\teff}s (by $\lesssim$ 100 K)
and [M/H] values (by $\lesssim$ 0.5 dex).

In the near infrared, spectral model fits again provided reasonably good
matches to the broad spectral energy distribution, with slight discrepancies
in the 0.99 $\micron$ FeH band, the 1.15--1.30 $\micron$ continuum (also
dominated by FeH absorption in late-type subdwarfs; Burgasser et al.\ 2004,
Cushing et al.\ 2006; Burgasser et al.\ in prep.)
and the 1.4 $\micron$ H$_2$O band.  In this wavelength regime,
Cond models typically provided the best fits, largely due to better
agreement in the 0.8--1.1 $\micron$ region.  Systematic shifts
in the derived parameters between the NextGen and Cond
model sets were less pronounced at these wavelengths, but higher surface
gravity models consistently gave larger {\teff}s (by 100--200 K)
and slightly lower [M/H] values (by $\lesssim$ 0.5 dex).  There was also
some degeneracy in the best fit values, 
with hotter, solar metallicity models providing 
fairly good matches to the data for $\lambda > 1.1 \micron$.  This is not
surprising, given that the blue, relatively featureless near infrared
spectra of esdMs are not unlike those of hotter M and K dwarf stars.

Derived parameters between the optical and near infrared fits for a given
object exhibit clear systematic differences.  {\teff}s derived from the
optical data are generally 100-200 K higher than those derived from
the near infrared data; metallicities are consistently 0.5--1.0 dex higher.
Examination of the models indicates that lower metallicities
are required to match the $H$- and $K$-band suppression in the near infrared
data, but can result in optical CaH bands that are too deep.
Similar effects are observed with lower {\teff}s.  Since we have
no way of independently validating fits in either spectral region,
we must treat these systematic differences as a source
of uncertainty in the model fits, emphasizing that absolute
{\teff}s and [M/H] values derived from spectral model fits 
should in general be treated with some caution.

Turning to the derived parameters of 
the objects in our sample, we find that, overall, later esdM types
correspond to cooler {\teff}s, consistent with expectations.  
The one notable exception is 
the esdM6.5 LSR 0822-1700,
for which we derive a temperature 100 K cooler than the esdM7 APMPM 0559-2903.
This slight difference may not be significant; indeed, our {\teff}
determination based on near infrared data is 200 K cooler than
that derived by \citet{lep0822} from optical data. However, 
metallicity effects may be at play. LSR 0822-1700 appears
to have a consistently lower [M/H] than APMPM 0559-2903 based on both optical  
\citep{lep0822} and near infrared analysis.  In any case, LEHPM 2-59
is the lowest temperature object in the group, with  
{\teff} $\approx$ 2800--3000 K. Evolutionary models by
\citet{bar97} predict a mass of $\sim$0.09~M$_{\sun}$ for these parameters
assuming an age of 10~Gyr, only $\sim$0.01~M$_{\sun}$ above the hydrogen
burning minimum mass for these metallicities.

\section{Discussion}

\subsection{The Temperature Scales of M Dwarfs and Extreme Subdwarfs}

The temperatures derived for late-type esdMs in this and other studies
\citep{sch99,leg00,lep0822} are relatively warm
compared to solar-metallicity dwarfs with comparable subtypes, which typically
have {\teff} $\sim$ 2200--2800 K \citep{kir93,leg98,leg00,gol04}.
Figure~\ref{fig7} compares {\teff} determinations for M dwarfs and esdMs
based on spectral model fits to NextGen models and their 
antecedents.  These data are segregated by
optical\footnote{We did not include {\teff} determinations for 
M dwarfs later than M6 in the \citet{kir93} and G97 studies due
to the poor quality fits to ultracool dwarf spectra existing models provided at that time.}
\citep{kir93,giz97,sch99,lep0822}
and near infrared \citep{kir93,leg96,leg00,leg01,daw00}
analyses. 
There is a significant downturn in the dwarf {\teff} scale around
type M5--M6, dropping from $\sim$3000 K to $\sim$2200 K by 
type M8.  The esdMs,
on the other hand, show a more gradual decrease in {\teff} with spectral
type over the same range.  Linear fits to optical and near infrared 
{\teff} determinations over the range esdM1-esdM8 yield parallel relations:
\begin{equation}
{\rm T}_{eff} = 3750 - 93{\times}{\rm SpT}_{esdM}~~{\rm K~(optical)} \label{eqn_esdmteff}
\end{equation}
\begin{equation}
{\rm T}_{eff} = 3620 - 96{\times}{\rm SpT}_{esdM}~~{\rm K~(near~infrared)} \label{eqn_esdmteff2}
\end{equation}
with dispersions of 30 and 90~K, respectively.  The temperatures of the two latest
esdMs, APMPM 0559-2903 and LEHPM 2-59, are 600--800~K hotter than 
those of equivalently typed solar-metallicity dwarfs.

Why would the effective temperatures 
of ultracool esdMs differ so greatly from those of ultracool 
solar metallicity dwarfs? It is well known that lower metallicity
implies lower overall atmospheric opacity,
a consequence of reduced molecular and
H$^-$ abundances.  Hence, for a given
mass, the observed photosphere lies at deeper, hotter and higher pressure
layers in more metal-poor stars.  However, this argument serves only
to explain (possible) differences
in {\teff}/mass relations between metallicity classes. {\teff}/spectral
type relations also hinge on how the spectral types themselves
are derived.  In fact, the classification of M dwarfs and
esdMs are determined by separate relations in the G97 and LRS03
schemes, as necessitated by the divergence of spectral
properties between these metallicity classes.  
Hence, differences in the {\teff}/spectral type
relations between M dwarfs and esdMs may simply be an artifact of the 
underlying classification scheme.  At first glance, this
may seem unimportant, for as long as the correct
scale is used for a given metallicity class, one should derive the correct
{\teff}.  However,  
metallicity is not a discrete quantity like metallicity class, 
and systematic deviations
can occur for individual objects that are more or less metal-poor than
the mean of their assigned class.  A more global {\teff}/metallicity/spectral
type relation is required.

\subsection{Future Prospects for Ultracool Subdwarf Classification}

With the recent compilation of several red optical proper motion surveys 
and the initiation of
near infrared programs \citep[Looper et al., in prep.]{dea05}, it is quite likely
that LEHPM 2-59's status as the coolest esdM known will not be
long-lived. Indeed, the identification of a few L subdwarfs (sdL) to date
suggests that esdL discoveries may not not far off (however, see below).  
It has been shown by \citet{me03f}
and \citet{lep1610} that these ultracool metal-poor stars and brown dwarfs
cannot be adequately classified by existing schemes largely
due to the disappearance of the 7150 {\AA} TiO band.
The lack of flux from these cool objects also 
argues against classification in the 6300-7200 {\AA} region.

So where do we proceed?  An obvious option is to define new subdwarf classification schemes at longer wavelengths, including the red optical
(cf.\ late-M and L dwarfs)
and near infrared (cf.\ T dwarfs)
spectral regions.  Absorption bands from the metal hydrides
FeH and CrH (and perhaps TiH; Burrows et al.\ 2005) are prominent in ultracool 
dwarf and subdwarf spectra for $\lambda >$ 8000 {\AA}, and are likely
to be present in even lower metallicity analogs.  Contrasting these
features with longer wavelength TiO (8200 and 8400 {\AA})
and VO bands (7900, 9500 and 10500 {\AA}), alkali lines
({\nai}, {\ki}, {\rbi} and {\csi}) or pseudocontinuum slope should prove to
be effective diagnostics for segregating metallicity and temperature
classes.  At longer wavelengths, subdwarf 
classification becomes more difficult
due to the weakness of absorption features and the strong suppression
of $H$- and $K$-band flux by H$_2$ absorption (which also wipes out the 
2.3 $\micron$ CO band).  Nevertheless, ratios comparing the
1.4 $\micron$ H$_2$O band and the near infrared spectral slope might
provide some discrimination (Burgasser et al.\ in prep.), while metal
line features (including {\ali};
see Cushing \& Vacca 2006) and FeH bands 
in the 1-1.35 $\micron$ region could be used for higher resolution studies.

As later extreme subdwarfs are identified, what will define the termination of the esdM class and the beginning of the esdLs?  Solar-metallicity L dwarfs
are distinguished by waning metal oxide bands, strengthening metal hydride
and alkali lines, steep red optical slopes and red near infrared colors.
Yet esdMs already exhibit weak metal oxides and strong metal hydrides, and
several neutral alkali species (including {\rbi} and {\csi}); while
near infrared colors will never become red due to strong absorption by 
H$_2$ \citep{sau94}.  Spectral models cannot provide clear guidance on this transition,
as it remains unclear as to whether condensate formation, a key aspect of the
M/L dwarf transition, plays a significant role in 
low-temperature metal-poor atmospheres \citep{me03f}.  Clearly,
cooler extreme subdwarf discoveries must guide the definition 
and our understanding of this transition.

One further complication in this issue is whether halo esdLs even
exist in our Galaxy today.  
As metal-poor halo objects about the substellar mass limit evolve
over $\sim$10~Gyr, wide gaps in the luminosities and effective
temperatures of this population develop.
Low mass stars attain a steady-state {\teff} $\gtrsim$ 2500-3000~K,
while most brown dwarfs cool
to {\teff} $\lesssim$ 1000~K in this time 
\citep[cf.\ Figure 3 in Burgasser et al.\ 2003a]{bur01}.  
Only a narrow range of low mass star/brown dwarf transition
objects are expected to
encompass intermediate L-type
{\teff}s between these limits, and
could therefore be quite rare.  Future surveys for low luminosity
metal-poor objects will hopefully probe this L dwarf gap, providing,
if little data for esdLs, a unique constraint
for brown dwarf evolutionary theories.

Finally, one area in which subdwarf classification can be immediately improved is the identification of specific standard stars for temperature
and metallicity subtypes.
A framework of standard stars is a fundamental tenet of the MK Process \citep{mor43,mor73,kee76,cor94},
the most widely adopted method of stellar classification.  Standard stars
provide the basis of current classification schemes for
solar metallicity M {\citep{kir91}, L \citep{kir99} and T dwarfs \citep{me06}.
Allowing specific stars to define a classification scheme
provides a level of consistency that is not generally present in pure
index schemes, while retaining independence from
constantly evolving theoretical interpretations.  There are now several
dozen M subdwarfs of types sdM5/esdM5 and earlier from which appropriate standards
can be chosen, and we anticipate that current red optical and near infrared
proper motion surveys will soon fill in the remainder of the M subdwarf and
extreme subdwarf sequences.  Now is an opportune time to
consider the construction of a more robust classification scheme for late-type subdwarfs.

\section{Summary}

We have identified LEHPM 2-59 as the coolest esdM identified to date.
Near infrared and optical spectroscopy show features indicative of a late-type
esdM, including strong alkali lines, VO and H$_2$O absorption
and a blue near infrared spectral slope.  We derive an optical 
spectral type of esdM8 on the G97 and
LRS03 schemes, and spectral model fits to both optical and
near infrared data indicate {\teff} = 2800--3000 K and 
-1.5 $\lesssim$ [M/H] $\lesssim$ -2.0 for this source.  
Its kinematics 
confirm it as a halo star.  We show that the temperatures of this and other
late-type esdMs are significantly higher than equivalently classified
solar-metallicity M dwarfs, a divergence that could lead to systematic
errors in derived parameters for objects with intermediate metallicities.
Finally, we have touched on the future prospects of M subdwarf
classification, and methods by which it may be improved and extended to
later subtypes.  
With the current high discovery rate of low-temperature,
ultracool halo subdwarfs, it is likely that these issues will require
further scrutiny in the near term.

\acknowledgments

We thank P.\ Hauschildt and F.\ Allard for making their
team's spectral models electronically available, and acknowledge useful
discussions with P.\ Hauschildt, S.\ L\'epine, S.\ Mohanty
and J.\ Mulchaey during
the preparation of the manuscript.
We also thank our anonymous referee and our scientific editor, J.\ Liebert,
for their helpful comments.
AJB acknowledges the assistance of telescope operators B.\ Golisch, D.\ Griep
and P.\ Sears at IRTF, and H.\ Rivera and
S.\ Vera at Magellan during the observations presented in this study,
as well as our instrument scientists J.\ Rayner (SpeX) and J.\ Bravo (LDSS-3).
This publication makes use of data 
from the Two Micron All Sky Survey, which is a
joint project of the University of Massachusetts and the Infrared
Processing and Analysis Center, and funded by the National
Aeronautics and Space Administration and the National Science
Foundation.
2MASS data were obtained from the NASA/IPAC Infrared
Science Archive, which is operated by the Jet Propulsion
Laboratory, California Institute of Technology, under contract
with the National Aeronautics and Space Administration.
The authors wish to extend special thanks to those of Hawaiian ancestry
on whose sacred mountain we are privileged to be guests.

Facilities: IRTF (SpeX); Magellan (LDSS-3)
%\facility{IRTF(SpeX)}; \facility{Magellan(LDSS-3)}

\clearpage

\begin{figure}
\centering
\epsscale{0.7}
%\plotone{f1.eps}
\caption{$H_J$ versus $(R_{ESO}-J)$ diagram for the LEHPM 
catalog \citep{por04}.  Note the three clusters of sources 
corresponding to (from center to lower left) main sequence dwarfs, 
halo subdwarfs and white dwarfs.  Our selection criteria for ultracool
dwarf and subdwarf
candidates are indicated by dashed lines.  Previously classified
sources within this region are noted by open circles; LEHPM 2-59
is indicated by an open square. [THIS FIGURE IS INCLUDED IN JPEG FORMAT]
\label{fig_rpm}}
\end{figure}

\begin{figure}
\centering
\epsscale{0.8}
\plotone{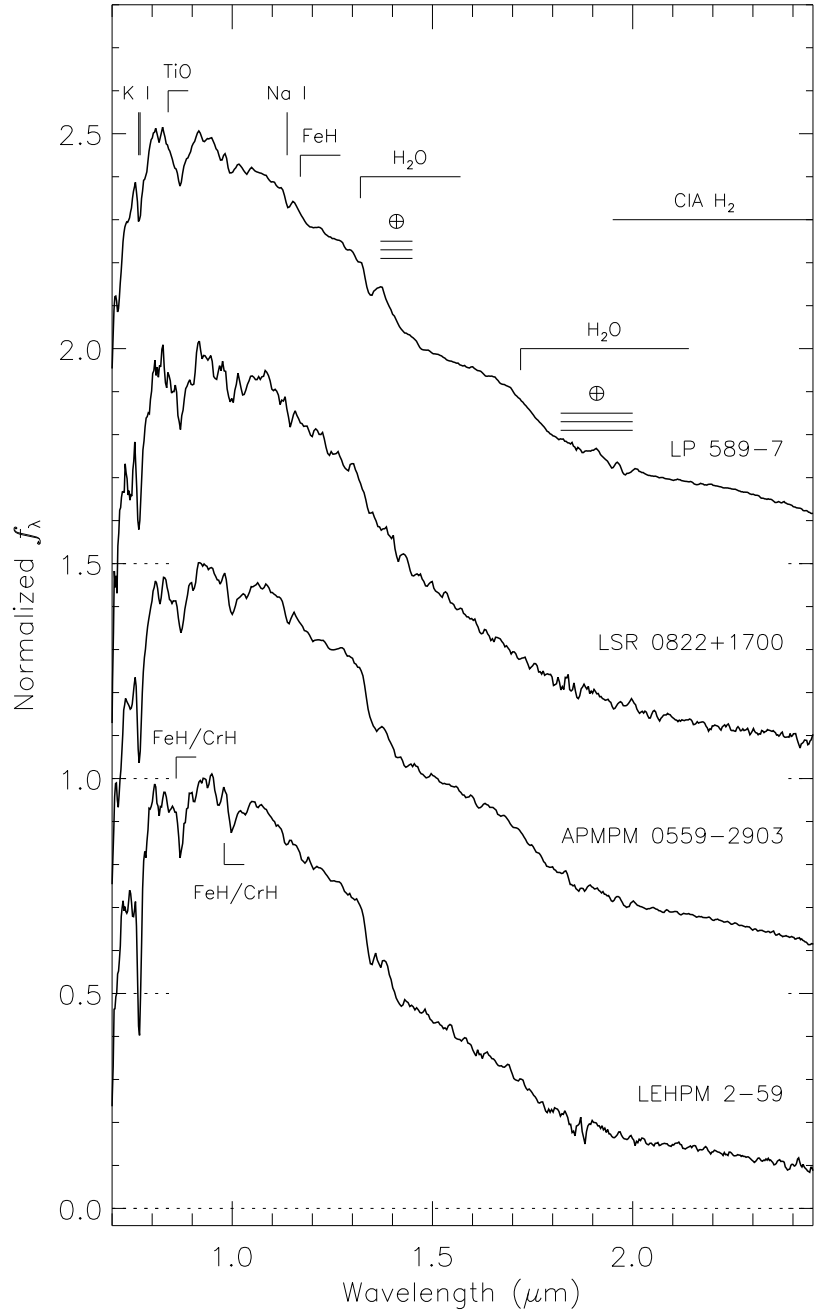}
\caption{SpeX prism spectra of the esdMs LP 589-7 (esdM5),
LSR 0822+1700 (esdM6.5), APMPM 0559-2903 (esdM7) and 
LEHPM 2-59 (esdM8).  Spectra are normalized at their flux peaks
and offset by constants (dotted lines).  Prominent features
in the NIR spectra of cool esdMs discernible at the resolution
of these prism data ({\ldl} $\approx$ 150) 
are labeled.  Regions of high telluric opacity
are indicated by $\oplus$ symbols.
\label{fig2}}
\end{figure}

\begin{figure}
\centering
\epsscale{0.8}
\plotone{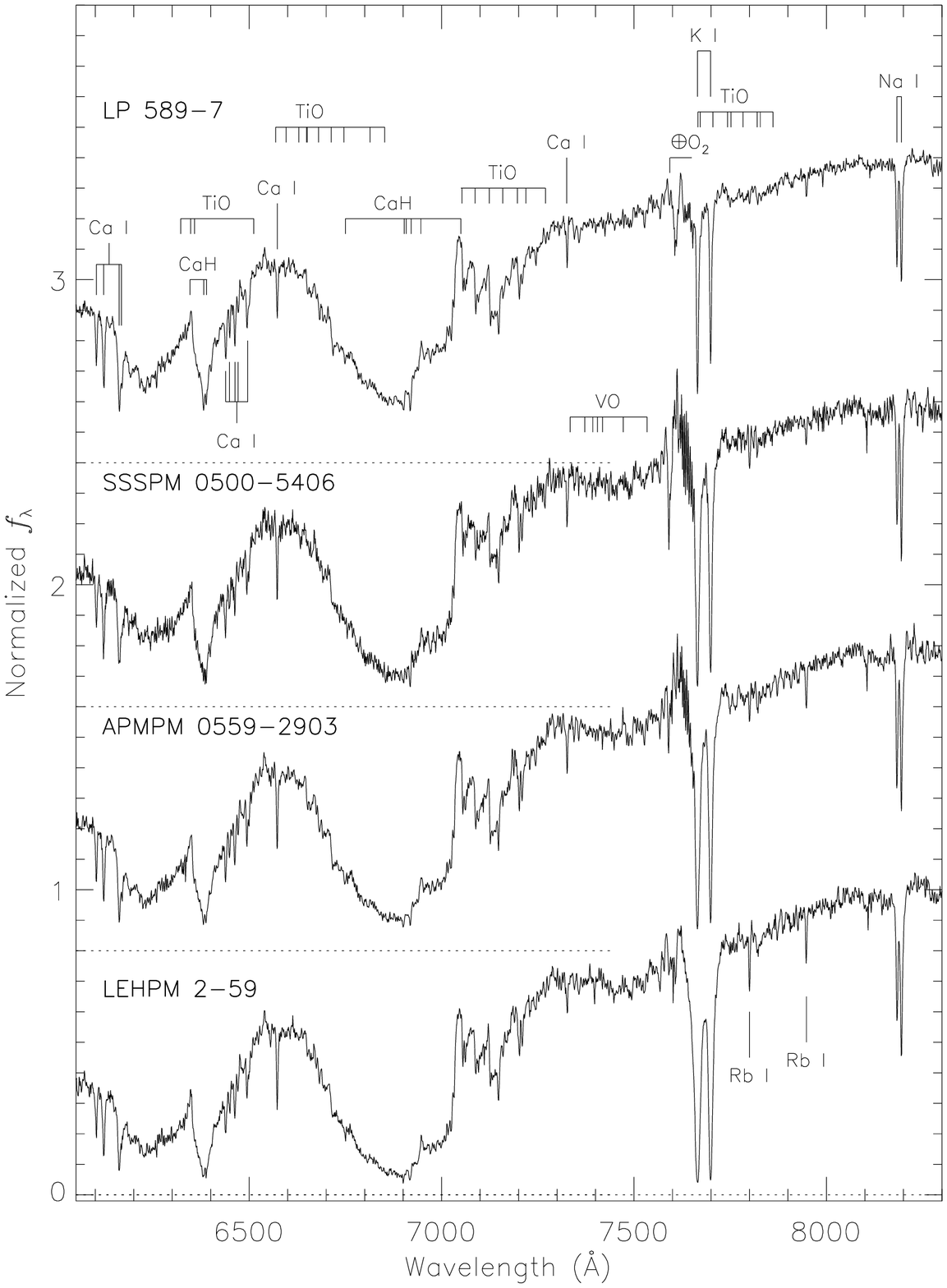}
\caption{Optical spectra of the late esdMs LP 589-7 (esdM5), SSSPM 0500-5406 (esdM6.5),
APMPM 0559-2903 (esdM7) and LEHPM 2-59 (esdM8) from top to bottom.  
All data are shifted to their rest frame velocities,
normalized at 8100 {\AA}
and offset by a constant (dotted lines).  Identified molecular features are from
\citet{kir91}; atomic features are from \citet[see Table 3]{kur95}. 
Residual noise from the telluric  O$_2$ A-band is indicated by 
$\oplus$ symbols.
\label{fig3}}
\end{figure}

\begin{figure}
\centering
\epsscale{0.8}
\plotone{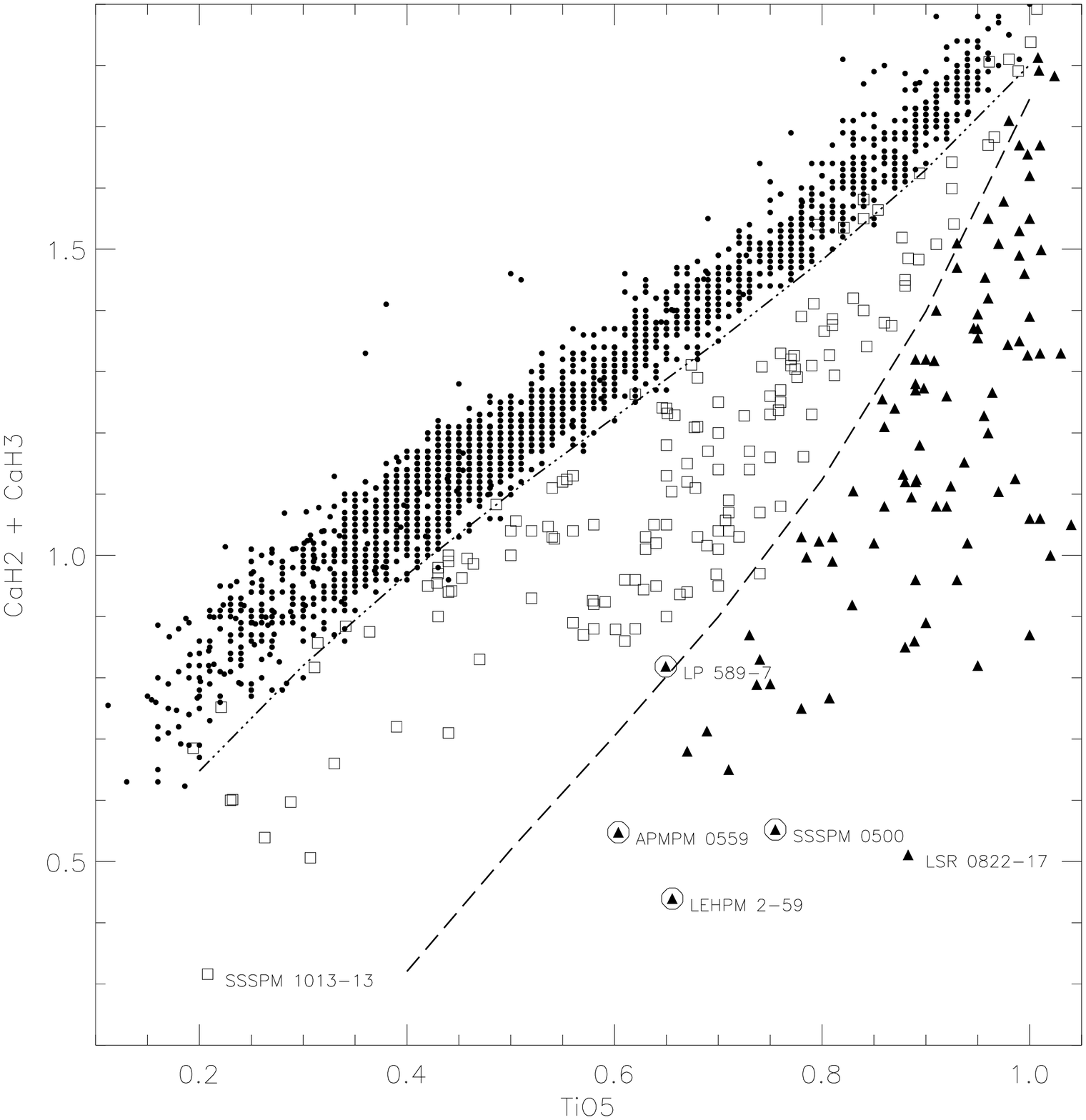}
\caption{Spectral indices CaH2+CaH3 versus TiO5 for dwarfs (points), 
subdwarfs (open squares) and extreme subdwarfs (filled triangles)
from \citet{haw96}; G97; \citet{giz97b};
\citet{rei02}; LRS03; \citet{lep0822}; and \citet{sch1013}.
Data from this paper
are encircled and labeled, while
values for the esdM6.5 LSR 0822+1700 and the sdM9.5 SSSPM 1013-1356 
from \citet{lep0822} and \citet{sch1013}, respectively, are also labeled.
Dashed lines delineate boundaries between dwarfs, sdMs and esdMs as defined by Eqns.~\ref{eqn_dsd} and~\ref{eqn_sdesd}.
\label{fig4}}
\end{figure}

\begin{figure}
\centering
\epsscale{1.0}
\plotone{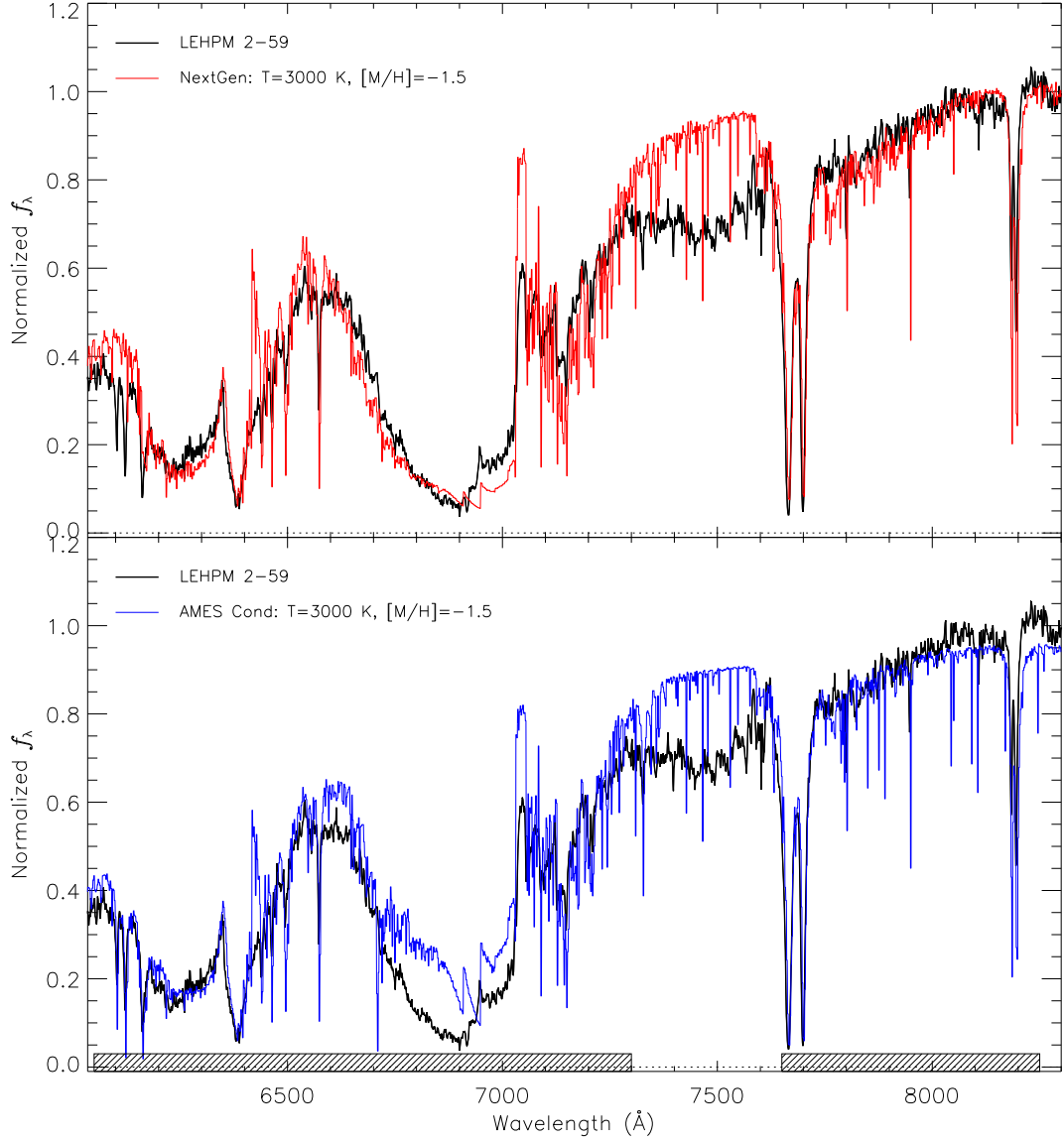}
\caption{Comparison of best fit NextGen (top) and AMES Cond (bottom)
$\log{g} = 5.0$ cm s$^{-1}$ models
to the observed red optical spectrum of LEHPM 2-59 (black).  LEHPM 2-59 
spectra are normalized at 8100 {\AA}; model spectra are scaled to their
best fit normalization.
The wavelength ranges for which spectral data and models were compared are
indicated by the hatched areas.
\label{fig5}}
\end{figure}

\begin{figure}
\centering
\epsscale{1.0}
\plotone{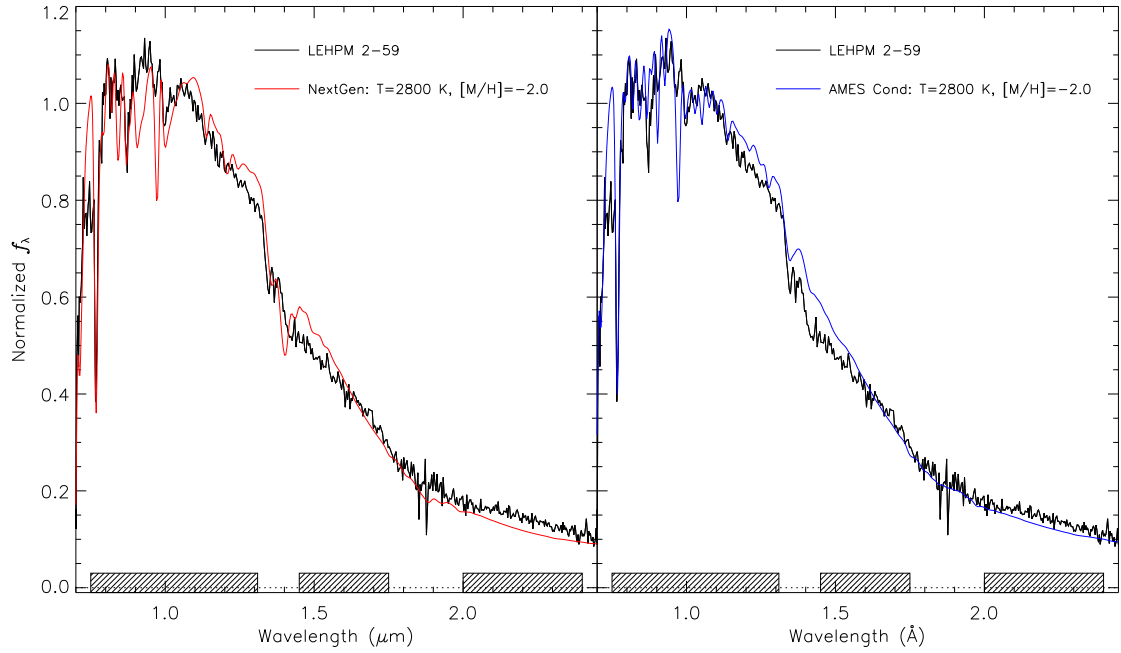}
\caption{Same as Fig.~\ref{fig5} for near infrared spectral fits to LEHPM 2-59. 
Spectra are normalized at 1.05 $\micron$.  
\label{fig6}}
\end{figure}

\begin{figure}
\centering
\epsscale{1.0}
\plotone{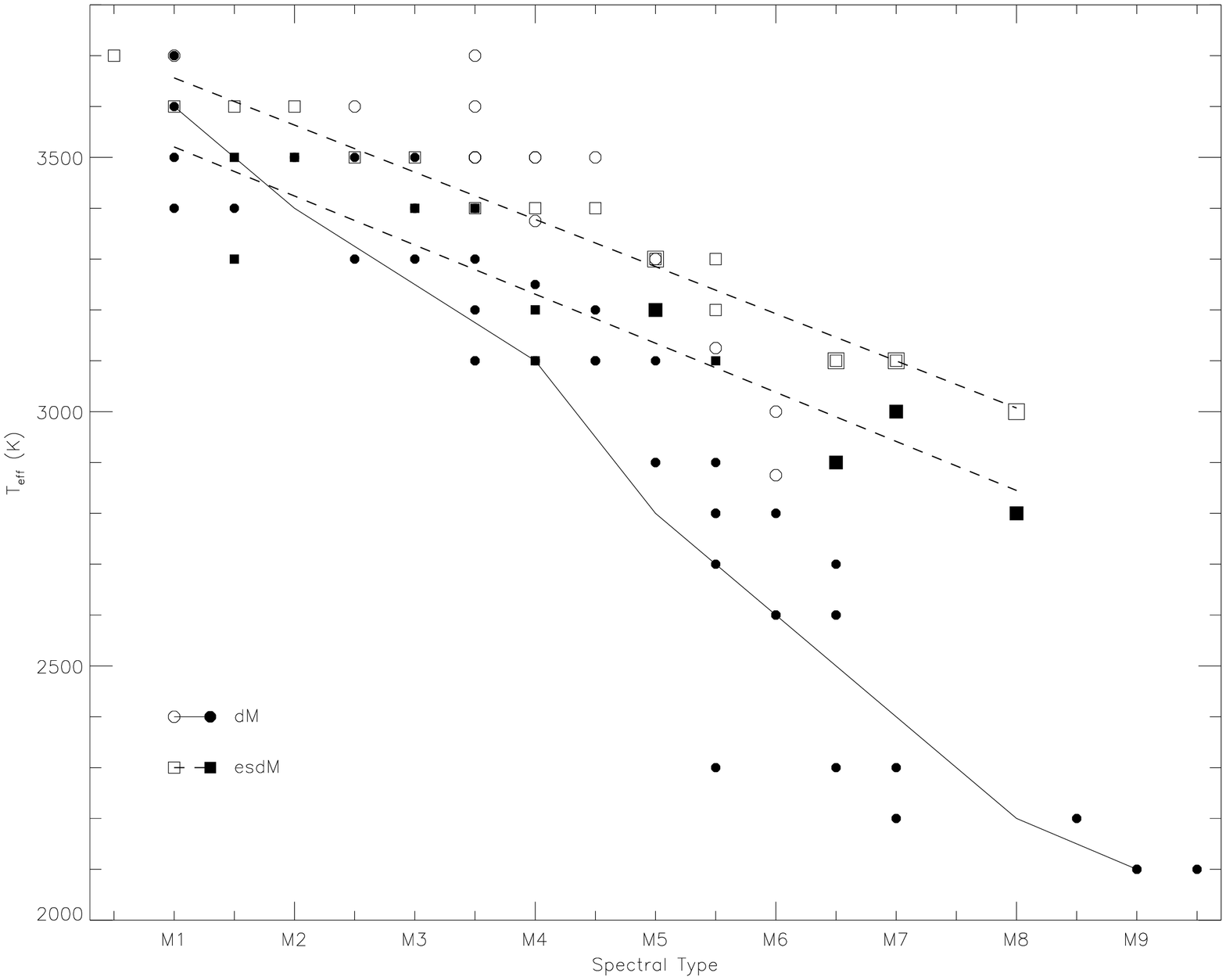}
\caption{{\teff}s for field M dwarfs (circles) and esdMs (squares) based on spectral 
model fits.  Data are from \citet[SpT $\leq$ M6]{kir93}; \citet{leg96,leg00,leg01}; 
\citet[SpT $\leq$ M6]{giz97}; 
\citet{sch99}; \citet{lep0822}; and this paper (oversize symbols).  {\teff}s derived from
fits to optical data are indicated by open symbols, those from near infrared data
by solid symbols.  The solid line delineates the M dwarf {\teff} scale
from \citet{rei00}; the dashed lines delineate linear fits for esdM
{\teff}s based on optical (top) and near infrared (bottom) data
(Eqns.~\ref{eqn_esdmteff} and~\ref{eqn_esdmteff2}).  
\label{fig7}}
\end{figure}

\clearpage

\begin{deluxetable}{lcccccclll}
\tabletypesize{\tiny}
\tablecaption{SpeX Observing Log \label{tab_spexlog}}
\tablewidth{0pt}
\tablehead{
\colhead{Source} &
\colhead{$\alpha$\tablenotemark{a}} &
\colhead{$\delta$\tablenotemark{a}} &
\colhead{$J$\tablenotemark{b}} &
\colhead{UT Date} &
\colhead{t$_{int}$ (s)} &
\colhead{Airmass} &
\colhead{Flux Cal.} &
\colhead{SpT} &
\colhead{Ref.\tablenotemark{c}} \\
}
\startdata
LP 589-7 &  01$^h$ 57$^m$ 27$\farcs$92 & +01$\degr$ 16$^m$ 43$\farcs$3  & 14.50$\pm$0.03 & 2004 Sep 5 & 1080 & 1.13 & HD 13936 & A0 V & 1 \\
LEHPM 2-59 &  04$^h$ 52$^m$ 09$\farcs$94 & -22$\degr$ 45$^m$ 08$\farcs$4 & 15.52$\pm$0.05 & 2004 Sep 9 & 720 & 1.43 & HD 32855 & A0 V &  2 \\
APMPM 0559-2903 &  05$^h$ 58$^m$ 58$\farcs$91 & -29$\degr$ 03$^m$ 26$\farcs$7  &  14.89$\pm$0.04 & 2005 Dec 31  & 1440 & 1.53 & HD 41473 & A0 V & 3 \\
LSR 0822-1700 &  08$^h$ 22$^m$ 33$\farcs$69 & -17$\degr$ 00$^m$ 19$\farcs$9  &  15.87$\pm$0.08 & 2004 Mar 10  & 1080 & 1.01 & HD 58383 & A0 V & 4 \\
\enddata
\tablenotetext{a}{J2000 Coordinates from 2MASS.}
\tablenotetext{b}{$J$ magnitudes from 2MASS.}
\tablenotetext{c}{Discovery reference for esdM source.}
\tablerefs{(1) \citet{giz99}; (2) \citet{por04}; (3) \citet{sch99}; (4) \citet{lep0822}}
\end{deluxetable}

\begin{deluxetable}{lcccccclll}
\tabletypesize{\tiny}
\tablecaption{LDSS-3 Observing Log \label{tab_ldss3log}}
\tablewidth{0pt}
\tablehead{
\colhead{Source} &
\colhead{$\alpha$\tablenotemark{a}} &
\colhead{$\delta$\tablenotemark{a}} &
\colhead{$R$\tablenotemark{b}} &
\colhead{UT Date} &
\colhead{t$_{int}$ (s)} &
\colhead{Airmass} &
\colhead{Tell.\ Cal.} &
\colhead{SpT} &
\colhead{Ref.\tablenotemark{c}} \\
}
\startdata
LP 589-7 &  01$^h$ 57$^m$ 27$\farcs$92 & +01$\degr$ 16$^m$ 43$\farcs$3  & 17.39 & 2005 Dec 4 & 2100 & 1.16 & HD 603 & G2 V & 1 \\
LEHPM 2-59 &  04$^h$ 52$^m$ 09$\farcs$94 & -22$\degr$ 45$^m$ 08$\farcs$4 & 18.72 & 2005 Dec 4 & 1800 & 1.04 & HD 31527 & G2 V & 2 \\
SSSPM 0500-5406 &  05$^h$ 00$^m$ 15$\farcs$77 & -54$\degr$ 06$^m$ 27$\farcs$3  & 17.42 & 2005 Dec 4 & 600 & 1.50 & HD 33967 & G2/3 V & 3 \\
APMPM 0559-2903 &  05$^h$ 58$^m$ 58$\farcs$91 & -29$\degr$ 03$^m$ 26$\farcs$7  &  18.08 & 2005 Dec 4 & 1200 & 1.06 & HD 33967 & G2/3 V & 4 \\
\enddata
\tablenotetext{a}{J2000 Coordinates from 2MASS.}
\tablenotetext{b}{Photographic $R$ magnitudes from the SuperCosmos Sky Survey.}
\tablenotetext{c}{Discovery reference for esdM source.}
\tablerefs{(1) \citet{giz99}; (2) \citet{por04};  (3) \citet{lod05}; (4) \citet{sch99}.}
\end{deluxetable}

\begin{deluxetable}{clccl}
\tabletypesize{\small}
\tablecaption{Atomic Lines Identified in esdM Optical Spectra over $\lambda\lambda$ 6100--8300 {\AA} \label{tab_atomic}}
\tablewidth{0pt}
\tablehead{
\colhead{Wavelength} &
\colhead{Element} &
\colhead{$E_{lower}$} &
\colhead{$E_{upper}$} &
\colhead{Ref.} \\
\colhead{({\AA})} &
\colhead{} &
\colhead{(eV)} &
\colhead{(eV)} &
\colhead{} \\
}
\startdata
6102.723 & \ion{Ca}{1}  & 1.879467  & 3.910663 &  1  \\
6122.217 & \ion{Ca}{1}  & 1.885935  & 3.910663 &  1  \\ 
6162.173  & \ion{Ca}{1} &  1.899063 &  3.910663 &  1   \\
6166.439\tablenotemark{a} & \ion{Ca}{1} &  2.521433 &  4.531641 &  1   \\
6169.042\tablenotemark{a} & \ion{Ca}{1} &  2.523157 &  4.532517  & 1   \\
6169.563\tablenotemark{a} & \ion{Ca}{1} &  2.525852 &  4.535043  & 1   \\
6439.075 & \ion{Ca}{1}  & 2.525852 &  4.450947 &  1   \\
6449.808 & \ion{Ca}{1}  & 2.521433 &  4.443325 &  1   \\
6462.567 & \ion{Ca}{1}  & 2.523157 &  4.441254 &  1   \\
6471.662 & \ion{Ca}{1}  & 2.525852 &  4.441254 &  1   \\
6493.781\tablenotemark{a} & \ion{Ca}{1} &  2.521433 &  4.430310  & 1   \\
6499.650\tablenotemark{a} & \ion{Ca}{1} &  2.523157 &  4.430310  & 1   \\
6572.779 &  \ion{Ca}{1} & 0.000000 &  1.885935  & 1 \\  
6717.681 & \ion{Ca}{1}  & 2.709192  & 4.554447 &  1   \\
7148.150 & \ion{Ca}{1}  &  2.709192 & 4.443325 &  2   \\
7202.200 & \ion{Ca}{1}  & 2.709192 &  4.430310 &  2   \\
7326.145 & \ion{Ca}{1}  & 2.932710 &  4.624710 &  2   \\
7664.911 &  \ion{K}{1}  & 0.000000 &   1.617220 & 1   \\
7698.974 &  \ion{K}{1}  & 0.000000 &   1.610064 &  1   \\
7800.259  & \ion{Rb}{1} &   0.000000  & 1.589158 & 3   \\
7947.597 & \ion{Rb}{1} &    0.000000  &  1.559697  & 3 \\
8183.255 &  \ion{Na}{1} & 2.102439 &   3.617221 &  4    \\
8194.790\tablenotemark{a} &  \ion{Na}{1} & 2.104571 &   3.617221 &  4    \\
8194.824\tablenotemark{a} &  \ion{Na}{1} & 2.104571 &   3.617215 &  4    \\
\enddata
\tablenotetext{a}{Blend.}
\tablerefs{(1) \citet{wie66}; (2) \citet{kur88}; (3) \citet{war68}; (4) \citet{kur75}.}
\end{deluxetable}

\clearpage

\begin{deluxetable}{lccccccc}
\tabletypesize{\scriptsize}
\tablecaption{Alkali Atomic Line Equivalent Widths ({\AA}).\label{tab_ews}}
\tablewidth{0pt}
\tablehead{
\colhead{Source} &
\colhead{{\ion{Ca}{1}}} &
\colhead{{\ion{Ca}{1}}} &
\colhead{{\ion{K}{1}}} &
\colhead{{\ion{Rb}{1}}} &
\colhead{{\ion{Rb}{1}}} &
\colhead{{\ion{Na}{1}}} &
\colhead{$V_{rad}$} \\
\colhead{} &
\colhead{(6573 {\AA})} &
\colhead{(7326 {\AA})} &
\colhead{(7665/7699 {\AA})} &
\colhead{(7800 {\AA})} &
\colhead{(7948 {\AA})} &
\colhead{(8183/8195 {\AA})} &
\colhead{(km s$^{-1}$)} \\
}
\startdata
LP 589-7 & 1.17$\pm$0.06 & 0.76$\pm$0.08 & 9$\pm$2 & $<$0.10 & 0.28$\pm$0.03 & 4.54$\pm$0.14 &    -82$\pm$9 \\
SSSPM 0500-5406 & 2.03$\pm$0.15 & 0.99$\pm$0.16 & 26$\pm$2 & $<$0.15 & 0.5$\pm$0.2 & 6.5$\pm$0.4 &  216$\pm$9 \\
APMPM 0559-2903 & 2.0$\pm$0.2 & 1.1$\pm$0.2 & 27.7$\pm$1.9 & 0.34$\pm$0.10 & 0.62$\pm$0.11 & 6.74$\pm$0.14 &  181$\pm$5 \\
LEHPM 2-59 & 2.17$\pm$0.13 & 0.93$\pm$0.12 & 39.7$\pm$1.2  & 0.60$\pm$0.07 & 0.68$\pm$0.05 & 7.2$\pm$0.3 & 79$\pm$9 \\
\enddata
\end{deluxetable}

\begin{deluxetable}{lcccccc}
\tabletypesize{\small}
\tablecaption{Spectral Ratios and Classification.\label{tab_class}}
\tablewidth{0pt}
\tablehead{
\colhead{Source} &
\colhead{CaH1} &
\colhead{CaH2\tablenotemark{a}} &
\colhead{CaH3\tablenotemark{a}} &
\colhead{TiO5} &
\colhead{Color-M\tablenotemark{a}} &
\colhead{SpT} \\
}
\startdata
LP 589-7 & 0.472 & 0.325 (4.8) & 0.493 (4.9) & 0.649 & 1.49 (4.7) & sd/esdM5 \\
SSSPM 0500-5406 & 0.310 & 0.220 (6.5) & 0.331 (7.0) & 0.755 & 1.64 (6.8) & esdM6.5 \\
APMPM 0559-2903 & 0.342 & 0.217 (6.6) & 0.331 (7.0) & 0.604 & 1.67 (7.1) & esdM7 \\
LEHPM 2-59 & 0.267 & 0.175 (7.3) & 0.265 (7.9) & 0.656 & 1.79 (8.7) & esdM8 \\
\enddata
\tablenotetext{a}{Numerical esdM subtype in parentheses based on Eqns.~1-3.}
\end{deluxetable}

\begin{deluxetable}{llcccccccc}
\tabletypesize{\scriptsize}
\tablecaption{Spectral Model Fits.\tablenotemark{a}\label{tab_model}}
\tablewidth{0pt}
\tablehead{
 & & \multicolumn{4}{c}{NextGen} & 
 \multicolumn{4}{c}{AMES Cond} \\
  & &  \multicolumn{2}{c}{5.0\tablenotemark{b}} &
 \multicolumn{2}{c}{5.5} & 
 \multicolumn{2}{c}{5.0} & 
 \multicolumn{2}{c}{5.5}  \\
 \cline{3-4} \cline{5-6} \cline{7-8} \cline{9-10}
\colhead{Source} &
\colhead{SpT} &
\colhead{\teff} &
\colhead{[M/H]} &
\colhead{\teff} &
\colhead{[M/H]} &
\colhead{\teff} &
\colhead{[M/H]} &
\colhead{\teff} &
\colhead{[M/H]} \\
\colhead{} &
\colhead{} &
\colhead{(K)} &
\colhead{(dex)} &
\colhead{(K)} &
\colhead{(dex)} &
\colhead{(K)} &
\colhead{(dex)} &
\colhead{(K)} &
\colhead{(dex)} \\
}
\startdata
\multicolumn{10}{c}{Optical} \\
\cline{1-10}
LP 589-7 & sd/esdM5 & {\bf 3300} & {\bf -1.0} & 3300 & -1.0 & 3200 & -1.0 & 3300 & -1.0  \\
SSSPM 0500-5406 & esdM6.5 & {\bf 3100} & {\bf -1.5} & 3200 & -1.5 & 3000 & -1.5 & 3200 & -1.0  \\
APMPM 0559-2903 & esdM7 & {\bf 3100} & {\bf -1.5} & 3200 & -1.0 & 3000 & -1.5 & 3100 & -1.0  \\
LEHPM 2-59  &  esdM8 & {\bf 3000} & {\bf -1.5} & 3100 & -1.5 & 3000 & -1.5 & 3100 & -1.0  \\
\cline{1-10}
\multicolumn{10}{c}{Near Infrared} \\
\cline{1-10}
LP 589-7 & sd/esdM5  & 3200 & -2.0 & 3300 & -1.5 & 3200 & -2.0 & {\bf 3200} & {\bf -1.5} \\
LSR 0822+1700 & esdM6.5  & 2900 & -2.0 & 3000 & -2.0 & {\bf 2900} & {\bf -2.0} & 3000 & -2.0 \\
APMPM 0559-2903 & esdM7 & 3000 & -2.0 & 3100 & -1.5 & 3000 & -1.5 & {\bf 3000} & {\bf -1.5}  \\
LEHPM 2-59 & esdM8   & 2800 & -2.0 & 3000 & -2.0 &  {\bf 2800} & {\bf -2.0} & 3000 & -2.0 \\
\enddata
\tablenotetext{a}{Values in bold denote overall best fit models.}
\tablenotetext{b}{{L}ogarithm of surface gravity in cm s$^{-2}$.}
\end{deluxetable}

\begin{deluxetable}{lcl}
\tabletypesize{\footnotesize}
\tablecaption{Properties of LEHPM 2-59.\label{tab_properties}}
\tablewidth{0pt}
\tablehead{
\colhead{Parameter} & 
\colhead{Value} &
\colhead{Ref.} \\
}
\startdata
${\alpha}$\tablenotemark{a} & 04$^h$ 52$^m$ 09$\farcs$94 & 1 \\
${\delta}$\tablenotemark{a} & -22$\degr$ 45$^m$ 08$\farcs$4 & 1 \\
SpT & esdM8 & 2,3,4 \\
%$B_J$\tablenotemark{b} & 21.72 & 2 \\
$R_{ESO}$\tablenotemark{b} & 18.72 & 5,6 \\
$R_{USNO}$\tablenotemark{b} & 18.82 & 7 \\
$I_{SSS}$\tablenotemark{b} & 16.86 & 5,6 \\
$I_{USNO}$\tablenotemark{b} & 17.24 & 7 \\
$I_{DENIS}$ & 16.74$\pm$0.09 & 8 \\
$J$ & 15.52$\pm$0.05 & 1 \\
$H$ & 15.25$\pm$0.08 & 1 \\
$K_s$ & 14.76$\pm$0.11 & 1 \\
$\mu$ & 0$\farcs$746$\pm$0$\farcs$016 yr$^{-1}$ & 6 \\
$\theta$ & $174{\fdg}7{\pm}1{\fdg}2$ & 6 \\
$d_{est}$ & 66$\pm$15 pc & 2,4 \\
$V_{rad}$ & $79{\pm}8$ km s$^{-1}$ & 2 \\
{[$U,V,W$]}\tablenotemark{c} & [135,-180,-80] km s$^{-1}$ & 2 \\
\teff & 2800-3000 K & 2 \\
{[M/H]} & -1.5 to -2.0 & 2 \\
\enddata
\tablenotetext{a}{2MASS coordinates, equinox J2000 and epoch 1998 Nov 29 (UT).}
\tablenotetext{b}{Photographic $R$ (IIIaF) and $I_N$ (IV-N) magnitudes.}
\tablenotetext{c}{Assuming a distance of 66 pc.}
\tablerefs{(1) 2MASS; (2) This paper; (3) \citet{giz97}; (4) \citet{lep03};
(5) SSS \citep{ham01a,ham01b,ham01c}; (6) \citet{por04}; (7) USNO-B1.0 \citep{mon03}; (8) DENIS \citep{epc97}.}
\end{deluxetable}

\end{document}